\documentclass{ws-procs975x65}

\usepackage{graphicx}
\usepackage{epsf}

\input epsf

\newcommand{\myskip}[1]{}

\renewcommand{\ni}{\noindent}

\renewcommand{\d}{{\rm d}}
\newcommand{\gs}{{\bar g}}

\newcommand{\mx}{m}
\newcommand{\BEQ}{\begin{eqnarray}}
\newcommand{\EEQ}{\end{eqnarray}}
\newcommand{\BEA}{\begin{eqnarray}}
\newcommand{\EEA}{\end{eqnarray}}

\newcommand{\Sigmab}{\overline\Sigma}

\newcommand{\Li}{{\rm Li}}

\newcommand{\kpc}{{\rm kpc}}

\newcommand{\eV}{{\rm eV}}

\begin{document}

\title{THE CASE OF 1.5 EV NEUTRINO HOT DARK MATTER}

\author{Theo M. Nieuwenhuizen}

\address{Institute for Theoretical Physics, University of Amsterdam\\
Valckenierstraat 65, 1018 XE Amsterdam, the Netherlands\\
$^*$E-mail: t.m.nieuwenhuizen@uva.nl\\http://staff.science.uva.nl/$\sim$nieuwenh}

\begin{abstract}
The lensing data of the galaxy cluster Abell 1689 can be explained by an
isothermal fermion model with a mass of 1-2 eV. The best candidate
is the 1.5 eV neutrino; its mass will be searched down to 0.2 eV in KATRIN 2015.
If its righthanded (sterile) modes were created too, there is 20\%
neutrino hot dark matter.
Their condensation on clusters explains the reionization of the intercluster
gas without Pop. III stars. Baryonic structure formation is achieved by
gravitional hydrodynamics alone, without dark matter trigger.
\end{abstract}

\bodymatter

\section{Introduction}
It is presently understood that the mass density of the universe is
(nearly) equal to the critical density, with a fraction
$\Omega_B\approx 4.5\%$ in baryons, $\Omega_D=20-25\%$
in dark matter and the rest in dark energy. In the standard view, dark matter is
cold, that is, a mass in the TeV regime, or at least warm, keV or more.
Well over 40 searches have been performed, which failed to detect the dark matter
particle, or came with claims that are not broadly accepted: the annual
variation of the signal recorded in DAMA or the ``two hints or background events'' of CDMS-II.

In Ref. \citen{Nneutrino09} the present author takes a ``blind'' approach
to describe the lensing data of the galaxy cluster Abell 1689. The assumptions are:
the dark matter is a non-interaction (quantum) gas;
the cluster is stationary;
it may be approximated as spherically symmetrical;
the mass distribution is isothermal for each of its components:
Galaxies, X-ray gas and dark matter;
all three components are subject to the common gravitational potential.
While galaxies and gas are so dilute that a classical isothermal model applies,
the dark matter may in principle be quantum degenerate, having
a Bose-Einstein or Fermi-Dirac distribution.
If the mass comes out large and thus the density low, either case will reduce to a
Maxwell-Boltzmann.

\section{Thermal fermion model}
The mass density of fermions with mass $m$, $\gs$ degrees of freedom
at a temperature $T$ in a gravitational
potential $U(r)$ and chemical potential $\mu=\alpha k_BT$ reads

\BEQ \label{rhoxp=} \rho_D=\!\int\!\frac{\d^3
p}{(2\pi\hbar)^3}\,\frac{\gs\mx }{\exp[(p^2/2\mx+mU(r)-\mu)/k_BT]
+1},\EEQ

\noindent
In dimensionless variables $x=r/R_\ast $, $\phi=mU/k_BT$, with
thermal length $\lambda_T=\left({2\pi\hbar^2}/{\mx k_BT}\right)^{1/2}$
and characteristic scale $R_\ast=({\lambda_T^3 k_BT}/{4\pi \gs G\mx^2})^{1/2}$,
the Poisson equation has
a polylogarithm Li$_\gamma(z)\!=
\!\sum_{k=1}^\infty \!z^k/k^\gamma$ for the dark matter component,
and Boltzmann terms for the galaxies (G)  and the X-ray gas (g),

\BEA \label{Poisson}
 \phi''+\frac{2}{x}\phi'=-\Li_{3/2}\left(- e^{\alpha-\phi}\right)
+ e^{\alpha_G-\bar\beta_G\phi}+e^{\alpha_g-\bar\beta_g\phi}.\EEA

This model is applied to lensing data of the galaxy cluster Abell 1689.
The plotted quantity is the 2D density contrast between the inner
disc with radius $r$ and the disc between $r$ and some fixed $r_m$,
  $\Delta\Sigmab(r)=[\Sigmab(r)-\Sigmab(r_m)]/{(1-r^2/r_m^2)}$.
Here $\overline\Sigma(r)=M_{2D}(r)/{\pi r^2}$ is the average projected mass profile,

\BEQ \label{Sig=A} \Sigmab(r)=
A\,\Phi\!\left(\frac{r}{R_\ast}\right),\quad \!\!
\Phi(x)=\int_0^\infty\!\d s\,\phi'(x\cosh s),\quad  A\equiv\frac{\hbar^6}{2\gs^2G^3m^8R_\ast^5}.\EEQ

\ni We make a $\chi^2$ fit of $\Delta\Sigmab$
to the 19 data points of Ref.~\citen{TysonFischer},
combined with 13 core points constructed from $M_{2D}(r)$ of Ref. ~\citen{Limousin}.
As seen in Fig. 1a, the relative errors increase strongly
with $r$, and become more equal if we consider the $\chi^2$ of
$\sqrt{\Delta\Sigmab}$.  We take~\cite{Nneutrino09}
$\bar\beta_g=0.153$ and $\alpha_g=2.36$ at $\bar\beta_G=1$. There is a
minimum $\chi^2=13.645$. With $h\equiv
0.70\,h_{70}$, the correlation matrix for the upper errors yields
$ A=59.4\pm9.6\,h_{70}M_\odot{\rm pc}^{-2},\quad
\alpha=38.4\pm 3.1,\quad
R_\ast=297\pm10 \,h_{70}^{-1}\kpc,\quad
\alpha_G=8.26\pm0.32.$
We present its fit in Figure 1a. The WIMP mass comes out as

\BEQ m=\frac{1}{2^{1/8}\gs^{1/4}}\,
\frac{\hbar^{3/4}}{G^{3/8}A^{1/8}R_\ast^{5/8}}=h_{70}^{1/2}\left(\frac{12}{\gs}\right)^{1/4}
1.455\pm0.030\,\,\eV. \EEA

\noindent This is much below the keV or TeV regime of the
supposed warm or cold dark matter, so the approach rules them out.
In Fig. 1b  we neglect the Galaxies and the gas, and plot the curves for
a Bose-Einstein distribution in the limit $\alpha\to 0^-$, for a classical isothermal model
and for an NFW profile~\cite{NFW} with $\bar\rho=\rho_c(1+z)^3$ at $z=0.183$:
$\rho_D=\bar\rho \delta_c r_s^3/[r(r+r_s)^2]$, $\delta_c=200 c^3/3[\ln(1+c)-c/(1+c)]$ with $c=4.684$.
None of them fits the data globally.

While $\gs$ is the number of states that can be filled
in the cluster formation process, $g$ is the filling factor in the
dark matter genesis. The global mass fraction thus reads

\BEQ\label{Omx=}
 \Omega_D=\frac{n_Fm}{\rho_c}=
\frac{g}{12}\left(\frac{12}{\gs}\right)^{1/4}h_{70}^{-3/2}
0.1893\pm 0.0039. \EEQ

\begin{figure}
\epsfxsize=14pc
\epsfbox{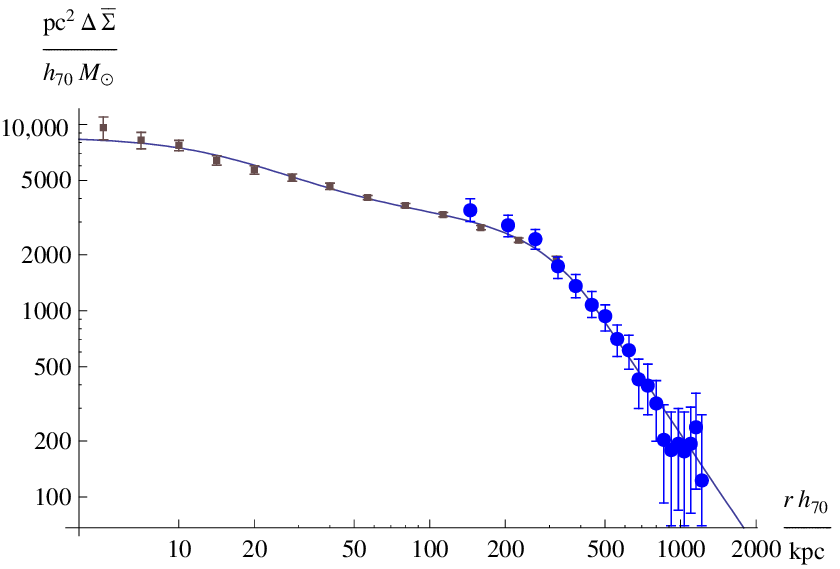}
\epsfxsize=14pc
\epsfbox{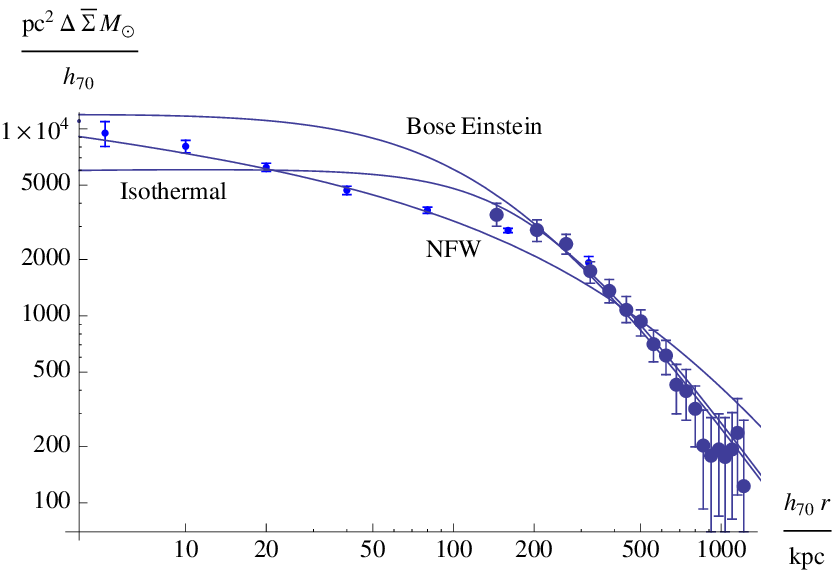}

\caption{ Left:
The mass contrast $\Delta \Sigmab$ as function of radius
$r$. Large data points from Ref.~\citen{TysonFischer}, small ones
(at radii $5\, 2^{n/2}h_{70}^{-1}$ kpc with $n=0,..,12$)
from Fig. 6 of Ref. \citen{Limousin}.
Full line: The theoretical profile.
Right:  1) Isothermal Bose-Einstein model with $\alpha\to0$;
2) Classical isothermal model; 3) NFW profile.
The first two do not fit for $r<200$ kpc, NFW not for $r>200-500$ kpc.}

\end{figure}

\section{ Neutrino: return of the first dark matter candidate}

They can occupy in the cluster formation
process all $\gs=12$ left and righthanded states, which gives
$m=1.445(30)\,h_{70}^{1/2}$ eV, below the Mainz-Troitsk bound of 2 eV. Neutrinos oscillate,
so their masses differ, but the effect is only in the meV regime.
The mass of the electron anti neutrino will be searched down to 0.2 eV in
the KATRIN tritium decay experiment (Karlsruhe, 2015).
So this will confirm or rule out our theory.
If the mass is indeed around 1.5 eV, it will have a 0.6 \% error,~\cite{Weinheimer}
so our approach will offer a new, sharp way to determine Hubble's constant.

With $g=6$ for active neutrinos, $\Omega_\nu\approx 10\%$. If righthanded
(and lefthanded anti-) neutrinos (sterile neutrinos) are created in the early
Universe, then $g\approx 12$ and $\Omega_\nu\approx 20\%$.
This may happen if there is a meV valued Majorana mass matrix~\cite{Cline}, that
by itself leads to neutrinoless double beta decay, be it a rather weak one,
the present upper bound is $\sim 0.5$ eV. With the Dirac mass matrix at the 1.5 eV
scale, the meV Majorana mass matrix presents an anti-see-saw mechanism.

Neutrinos are free streaming at the decoupling and condense on the galaxy cluster
fairly late, at $z\sim7$. The theory of violent relaxation predicts a Fermi-Dirac
distribution. 
Since the cosmic baryonic voids are till then filled with neutrinos,
the metric is then quite homogeneous. It is intriguing to see whether
the present inhomogeneity can explain the cosmological constant of the
concordance model.

The region in Abell 1689 where the neutrinos are quantum degenerate
 is millions of light years wide, a truly large scale for quantum behavior.

Gravitational hydrodynamics explains baryonic structure formation without cold dark matter trigger,
because viscosity is more important than often realized~\cite{NGS09,NMG12nu}
Galactic dark matter is observed as MACHOs of earth weight
and in other ways.~\cite{NGS09,NMG12nu}
Therefore neutrino free streaming is not ruled out by structure formation arguments.


\begin{thebibliography}{9}
\bibitem{Nneutrino09} Th. M. {Nieuwenhuizen},{Europhys. Lett.}
{\bf 86}, {59001} (2009).


\bibitem{TysonFischer} J. A. Tyson and P. Fischer, Astroph. J.
{\bf 446}, L55 (1995).


\bibitem{Limousin} M. Limousin et al., Astroph. J. {\bf 668},
643 (2007).

\bibitem{Cline} J. M. Cline, Phys. Rev. Lett. {\bf 68}, 3137 (1992).

\bibitem{NFW}
J. F. Navarro, C. S. Frenk and S. D. M. White,
Astrophys. J. {\bf 490}, 493 (1997).

\bibitem{Weinheimer} C. Weinheimer, http://arxiv.org/pdf/0912.4612.

\bibitem{NGS09}
Th. M. Nieuwenhuizen, C.H. Gibson and R. E. Schild,
Europhys. Lett. {\bf 88}, 49001 (2009).

\bibitem{NMG12nu} 
Th. M. Nieuwenhuizen, C.H. Gibson \& R. E. Schild,
 arXiv:1003.0453;  these proceedings.


\end{thebibliography}
\end{document}